\documentclass[usenatbib,usegraphicx,referee]{mn2e}

\def\P{{\cal P}}

\def\th{\vec{\theta}}
\def\NH{N_{\rm HI}}
\def\HI{{\rm H~\sc{i }}}
\def\x{\vec{x}}
\def\k{\vec{k}}
\def\u{\vec{U}}
\def\dI{\tilde{I}}
\def\dNH{ \tilde{N}_{\rm HI}}

\usepackage{epsfig}
\begin{document}
\title{The \HI column density power spectrum of six nearby spiral galaxies} 
\author[Prasun Dutta, Somnath Bharadwaj]
{Prasun Dutta$^{1}$\thanks{Email:prasun@ncra.tifr.res.in},  
and Somnath Bharadwaj$^{2}$\thanks{Email: somnath@cts.iitkgp.ernet.in}
\\$^{1}$ National Centre For Radio Astrophysics, Post Bag 3,
Ganeshkhind, Pune 411 007, India.
\\$^{2}$Department of Physics and Meteorology \& Center for
Theoretical Studies, Indian Institute of 
Technology, Kharagpur, Pin 721302, India.}

\date{}
\maketitle
\begin{abstract}
We  propose a method to determine the power spectrum of 
\HI column density  fluctuations using   radio-interferometric
observations of  21-cm 
emission from the ISM of galaxies.  We have used this to 
estimate the power spectra  of six  nearly face on   nearby spiral
galaxies.  Earlier work has shown 
that these power spectra are well fitted  by power laws with slopes 
around $-1.6$  across length-scales $\sim 1 \, {\rm kpc}$ to $\sim 10 \, {\rm
  kpc}$, the amplitude however was undetermined.  In the present work
we have determined the amplitude of  the \HI column density power
spectrum.   We find that  the \HI column density  $N_{\rm   HI}$ 
expressed in units of $10^{20} \, {\rm cm}^{-2}$ has mean square
fluctuations in the range $\sim 0.03$ to 
$\sim 20$.   The amplitude of the power
spectrum is found to be tightly correlated with the \HI mass 
fraction of the galaxies. The physical process responsible for these
scale-invariant fluctuations is, however,  at present not known. 
\end{abstract}
\begin{keywords}
physical data and process: turbulence-galaxy:disc-galaxies:ISM
\end{keywords}

\section{Introduction}
The statistical distribution of the neutral hydrogen (\HI)  column
density  is  a  major topic of research  in cosmology 
(see \citealt{2009RvMP...81.1405M} for a review).
The column density distribution  $f(N_{HI}, z)$ is well 
studied at   high redshifts using quasar absorption spectra.  
This has also been studied at $z=0$ using high resolution \HI 21-cm
emission maps of nearby galaxies (e.g., \citealt{2005MNRAS.364.1467Z}). 
Measurements of the \HI column density distribution span 
from $N_{HI} \sim  10^{12}\,  - \,  1.6 \times 10^{17} {\rm cm}^{-2}$ 
which  corresponds to  the highly ionized gas in 
the intergalactic medium (IGM)  (e.g. \citealt{2000ApJ...543..552S}) 
to   $N_{HI} > 10^{19} {\rm cm}^{-2}$ which is  the 
 predominantly  neutral gas in
the inter-stellar medium (ISM) of protogalaxies and  galaxies 
(e.g., \citealt{2010ApJ...718..392P}).  Observations of the \HI column
density distribution encompass a large variety of astrophysical
systems, and  this is an important probe  of the 
CDM structure formation paradigm.

In this work we focus on the ISM of nearby galaxies which corresponds
to the high column density  regime ($N_{HI} > 10^{19} {\rm
  cm}^{-2}$). The high redshift counterparts of such systems appear as
the  damped Lyman $\alpha$  (DLA; $N_{HI} > 2 \times 10^{20} \, {\rm
  cm}^{-2}$)   systems  (e.g. \citealt{2005ARA&A..43..861W}). The
column 
density distribution  $f(N_{HI},z)$ shows two interesting
features in this regime. First, a steepening is observed at 
 $\NH \ge  10^{21} \, {\rm cm}^{-2}$, and then 
there is a sharp decline  for  $\NH \ge  10^{22} \ {\rm  cm}^{-2}$
where molecular hydrogen  start to form. 
\citet{2012ApJ...761...54E} have shown that the turnover at $\sim
10^{21} \, {\rm cm}^{-2}$  is an effect of  the  orientations of
early spiral galaxies.  The same turnover occurs at a lower column
density \citep{2013MNRAS.429.1596P} for the nearby dwarf galaxies
owing to their spheroidal geometry \citep{2010MNRAS.404L..60R}.   

While there has been a large amount of  work towards observing and 
understanding the \HI column density distribution, it has been    mainly 
restricted to the distribution function $f(N_{HI},z)$ which  is a
one-point statistics.  This  does not quantify the spatial clustering
of the \HI column density distribution, and it is necessary to use two
point and higher statistics for this.  Here  we use the power
spectrum  which is a two point statistics, it being 
the Fourier transform of the two-point correlation function.
In this work   we use  the \HI column density power spectrum 
$P_{\NH}(k)$  to quantify the clustering of the fluctuations in the \HI 
column density as a function of the two dimensional (2D) wave number
$k$ or equivalently the length-scale $2~\pi/k$. 

Several earlier studies have used   radio-interferometric observations 
 to   measure the  specific intensity  power spectrum 
 of \HI  21-cm emission from
 the ISM of our Galaxy and external galaxies. 
\citet{1983A&A...122..282C} and  \citet{1993MNRAS.262..327G} 
find  power-law  behavior with
slopes in the range   $ -2.5$ to $-2.8$ at length scales ranging from 
 $ 10$ pc to $200$ pc within our Galaxy. 
The  Small Magellanic  Cloud   \citep{1999MNRAS.302..417S}, 
the Large Magellanic Cloud  \citep{2001ApJ...548..749E}
and the dwarf galaxy DDO~210 \citep{2006MNRAS.372L..33B} all show
power law behavior  similar to  our Galaxy.  Subsequent studies  
\citep{2008MNRAS.384L..34D, 2009MNRAS.397L..60D, 2009MNRAS.398..887D,
  2013NewA...19...89D} have analyzed  several dwarf and spiral
galaxies to clearly establish a power law behavior to length scales
as large as  $10 \, {\rm kpc}$. The slope, however, was found to be
around $-1.6$ at large scales ($1 \, - \, 10 \, {\rm 
  kpc}$)  in contrast to the value  of around $-2.5$ found at smaller
scales ($10 \, - \, 200 \, {\rm   pc}$).  This has been interpreted as
a transition from two-dimensional (2D) turbulence  in the plane of the
galaxy's disk at large scales   to three dimensional (3D) turbulence
at length-scales smaller than the scale-height of the disk. 
 
The 21-cm emission is directly proportional to the \HI column
density. However, it is not straight forward to extract the \HI column
density power spectrum directly from the visibilities measured in 
radio-interferometric observations. Consequently, the earlier works
have all focused on the slope of the power spectrum. The amplitude of
the \HI column density power spectrum is, at present, largely unknown. 
It has been possible to measure  both the amplitude and slope of the \HI
optical depth  power spectrum  at relatively  small scales ($0.01 \,$
to $3 \, {\rm pc}$; \citealt{2000ApJ...543..227D, 2010MNRAS.404L..45R}
). However, it is necessary to assume the value of the spin
temperature in order to convert this to the  \HI column density power
spectrum. Further, the  the \HI  column
density and  the spin temperature both contribute
to the \HI optical  depth. This  makes it difficult to interpret the \HI
optical depth  in terms of the \HI column density power
spectrum.  

 In this {\em Letter} we propose a method to estimate the \HI column density
 power spectrum using  radio-interferometric observations of the 21-cm 
emission from the ISM of galaxies. We apply the method  to a sample of six 
external spiral galaxies drawn  from The \HI
Nearby Galaxy Survey (THINGS; \citealt{2008AJ....136.2563W}). The
galaxies that we have analyzed  are  
nearly face-on,  and  we do not expect geometrical effects
to be important in  this study.  The method is presented in Section~2,
while Section~3 presents the data and results, and we have the
discussion and conclusion in Section~4. 

\section{Method}
{\bf The specific intensity of  21 cm emission $I_{\rm HI}(\th, \nu)$
observed at a frequency $\nu$ and direction $\th$ on the sky 
is related to the \HI 
 column density $\NH(\th)$  as
\citep{2011piim.book.....D}}
\begin{equation}
I_{\rm HI}(\th, \nu)=\left( \frac{3 A_{21} h \nu_e}{16 \pi} \right)
\phi(\nu) \NH(\th), 
\label{eq:nh1}
\end{equation} 
where $\nu_{e}$ is the rest frame frequency of the \HI emission,
$A_{21}$ is the Einstein coefficient and $\phi(\nu)$ is 
the line shape function with $\int \phi(\nu) d\nu 
= 1$. Note that  the optical depth is assumed to be
small compared to unity in deriving the above equation. We shall
discuss  this assumption later. 
Above eqn.~(\ref{eq:nh1}) may also be written as 
 \begin{equation}
I_{\rm HI}(\th, \nu)= C_1 \phi(\nu) \NH(\th) \,,
\label{eq:nh2}
\end{equation}
where 
$C_1=1.62 \times 10^{5} \, {\rm Jy \, sr^{-1} \, Hz}$ 
and in eqn.~(\ref{eq:nh2}) and everywhere  subsequent to this we express 
$\NH$ in units of  $10^{20} \,
{\rm  \, cm}^{-2}$.  \\
First, we collapse all the frequency channels with \HI emission, 
{\it ie.}
\begin{equation}
I_{\rm HI}(\th)=\int \, I_{\rm HI}(\th, \nu) \, d \nu =C_1  \NH(\th).
\end{equation}
It is convenient to express $\NH$ as a function
of the two-dimensional position vector $\x=r \th $ defined at the
distance $r$ to the location of the \HI.  {\bf For an
 external galaxy  $r$ refers to the distance to the galaxy.} 
 The respective Fourier
transforms of $I_{\rm HI}(\th)$ and $ \NH(\x)$ are defined as 
\begin{equation}
\dI(\u)=\int d^2 \theta \, e^{2 \pi i \th \cdot \u} \, 
I_{\rm HI}(\th)
\label{eq:nh3}
\end{equation}
and 
\begin{equation}
\dNH(\k)=\int d^2 x \, e^{i \k \cdot \x} \, 
\NH(\x),
\label{eq:nh4}
\end{equation}
where $U$ is the inverse angular scale, {\bf $2 \pi U$ is the angular wave
number  and $k$ is the 2D spatial wave number.} 
Comparing eqn.~(\ref{eq:nh3}) and eqn.~(\ref{eq:nh4}) we see that 
$\k=2 \pi \u/r$ and 
\begin{equation}
\dI(\u)=\left[\frac{C_1}{r^2} \right] \dNH(\k).
\label{eq:nh5}
\end{equation}
We define the power spectra 
\begin{equation}
\langle \dI(\u) \dI^*(\u^{'}) \rangle = \delta^2_D(\u -\u^{'}) \,
P_{\rm HI}(U)
\label{eq:nh6}
\end{equation} 
and 
\begin{equation}
\langle \dNH(\k) \dNH^*(\k^{'}) \rangle =(2 \pi)^2 \delta^2_D(\k -\k^{'}) \,
P_{\NH}(k)
\label{eq:nh7}
\end{equation}
where  {\bf $\delta^2_D(\u)$ is the 2D Dirac delta function },  $P_{\rm
  HI}(U)$  is the  angular power spectrum  of the  
specific intensity of the \HI 21-cm  emission  and $P_{\NH}(k)$ is  
the two-dimensional {\bf spatial}  power spectrum of the  \HI 
column density distribution. 
We also have 
\begin{equation}
 \delta^2_D(\u)= \left(\frac{2 \pi}{r}\right)^2  \delta^2_D(\k)
\end{equation}
whereby 
\begin{equation}
P_{\rm HI}(\u)=\left[ \frac{C_1}{r} \right]^2 P_{\NH}(\k).
\label{eq:nh8}
\end{equation}
In radio-interferometric observations an external galaxy  usually
occupies only a small fraction  of the telescope's  
field of view. Hence, we may neglect the effect of the telescopes
aperture here. If the galaxy's \HI disk  subtends a solid angle
$\Omega_{g}$ on the sky, we may write
\begin{equation}
I(\th) = W(\th)\, I_{\rm HI}(\th),
\end{equation}
where we define $W(\theta)$ such that $W(\theta) = 1$ inside the
galaxy and $0$ outside with $\int W(\th) d \th =
\Omega_{g}$. We define the  extent of the galaxy's \HI disk  using the
criterion that  the column density should exceed a cut-off
value $(N_{\rm HI}(\th) >N_{\rm HI}^{c})$ within the galaxy 
 and  regions with lower column densities  are excluded. We discuss
 our choice of  $N_{\rm HI}^{c}$ later.   
The measured visibilities  are collapsed across the frequency channels
with \HI emission, {\it ie.},
\begin{equation}
V(\u)=\Delta \nu_c \, \sum_{i} V(\u, \nu_i),
\end{equation}
where $\Delta \nu_c$ is the channel width in Hz. The resulting
frequency collapsed visibility  can be written as
\begin{equation}
V(\u) = \tilde{W} (\u) \otimes \tilde{I}_{\rm HI} (\u),
\end{equation}
where $\tilde{W} (\u)$ is the Fourier transform of $W(\th)$.
We then have 
\begin{equation}
\langle \mid V(\u, \nu) \mid^{2} \rangle = \mid \tilde{W} (\u) \mid^{2}  \otimes 
P_{\rm HI}(\vec{U}).
\end{equation}
For large baselines ($U^2 \Omega_g \gg 1$) we nay approximate this as 
\begin{equation}
\langle \mid V(\u, \nu) \mid^{2} \rangle = \Omega_{g}\, P_{\rm HI}
(\vec{U}).
\end{equation}
whereby 
\begin{equation}
k^2\, P_{N_{\rm HI}}(\k)\ =\ \left [\frac{2 \pi}{C_{1}} \right]^{2}\, 
\frac{\langle \mid V(\u) \mid^{2} \rangle\, U^{2}}{\Omega_{g}} \, .
\end{equation}
The dimensionless quantity $k^2 P_{N_{\rm HI}}(\k)$  directly
gives an estimate of the mean square fluctuation of the \HI column
density at the length-scale $2 \pi/k$. \\
Here we focus on a situation where  the visibility correlation 
is well fitted by a power law 
\begin{equation}
\langle \mid V(\u, \nu) \mid^{2} \rangle = A_{{\rm HI}}\,
\left[\frac{U}{{\rm k} \lambda}\right]^{\alpha},
\label{eq:pw1}
\end{equation}
where we may write
\begin{equation}
k^2\, P_{N_{\rm HI}}(\k)\ =\ A_{{\rm N}_{\rm HI}} \left[\frac{k}{{\rm
      kpc}^{-1}} \right]^{\alpha+2}, 
\label{eq:amp}
\end{equation}
with 
\begin{equation}
 A_{{\rm N}_{\rm HI}}\ =\ \left [\frac{2 \pi}{C_{1}}
  \right]^{2}\, \frac{1}{\Omega_{g}}\, \left( \frac{r}{2 \pi \ {\rm
    Mpc}} \right)^{\alpha+2}\ A_{{\rm HI}}.
\label{eq:ampform}
\end{equation}
{\bf Here $A_{{\rm N}_{\rm HI}}$  gives  the amplitude of the \HI  
column density power spectrum at the wave number $k= 1 \, {\rm
  kpc}^{-1}$  or equivalently the length scale  $2\, \pi \, {\rm
  kpc}$.} 

{\bf To summarize this section, radio interferometric observations  provide a direct estimate  of the angular
power spectrum of the  source specific intensity  distribution.  With the knowledge of the 
distance and a correction for  the source size,  this gives an
estimate of  the 2D spatial power spectrum  of the \HI column density
distribution.}

\section{Data and Results}
\citet{2008AJ....136.2563W} reports Very Large
Array \footnote{National Radio Astronomical Observatories, Very Large
  Array (NRAO VLA)} observations of a sample of 34 nearby 
galaxies as a part of The \HI Nearby Galaxy
Survey (THINGS). These observations have high angular ($\sim 6^{''}$ ) and
velocity ($\le 5.2$ km s$^{-1}$)
resolution. \citet{2013NewA...19...89D}  have used  visibility
correlations to 
estimate the \HI 21-cm emission specific intensity power spectra $P_{\rm
  HI}(U)$ of $18$ galaxies in the THINGS sample.  The 
details of the visibility correlation estimator  can be
found in \citet{2006MNRAS.372L..33B},  and
\citet{2009MNRAS.398..887D}, and we do 
not discuss it here. In addition to the requirement that the  galaxy's
minor axis should be greater than $6^{'}$
\citep{2013NewA...19...89D},  for the present study we also impose the
requirement that  average  inclination angle should be less than
$35^{\circ}$.   
This is because geometrical projection effects becomes important 
at higher inclination angles for which it is incorrect to assume that
the optical depth is directly proportional to the \HI 21-cm specific
intensity. This restricts our analysis to a sub-sample $6$  galaxies  
from  THINGS.

Table~1 summarizes different properties
of the six galaxies that  we have analyzed here. The  columns in
Table~1  are 
as follows: (1) name of the galaxy, (2) 
 distance to the galaxy, and (3) the average \HI inclination
 angle. The distances to the galaxies are  from 
\citet{2008AJ....136.2563W}, whereas the inclination angles are
from \citet{2008AJ....136.2648D}. \citet{2013NewA...19...89D} has
determined the $U$ range  ($U_{min}$ to $U_{max}$)  across which it is
possible to fit a power law to the measured  $P_{\rm HI}(U)$, 
columns  (4) and (5) of Table~1 gives the best-fit  values for
$A_{{\rm HI}}$ and  $\alpha$.  

\begin{table}
\centering
\begin{tabular}{l r r r r }
\hline
Galaxy & r & $i$ & $A_{{\rm HI}}$ & $\alpha$ \\
 & (Mpc) & ($^{\circ}$) & ($\times 10^{2}$)& \\ 
\hline \hline
NGC~628   &	$7.3$    &    $15.0$  &	$2.2\pm0.3$	& $-1.6\pm0.1$ \\
NGC~3184  &	$11.1$   &    $29.0$  &	$0.6\pm0.1$	& $-1.3\pm0.1$ \\
NGC~5194  &	$8.0$    &    $30.0$  &	$4.6\pm0.3$	& $-1.7\pm0.1$ \\
NGC~5457  &	$7.4$    &    $30.0$  &	$45\pm2$	& $-2.1\pm0.1$ \\
NGC~5236  &	$4.5$    &    $31.0$  &	$15\pm3$	& $-1.8\pm0.1$ \\
NGC~6946  &	$5.9$    &    $35.0$  &	$0.6\pm0.1$	& $-1.7\pm0.1$ \\
\hline
\end{tabular}
\label{tab:sample}
\caption{This shows  different parameters for the six galaxies that we
  have 
  analyzed here.  Columns (2) to (5) give the  distance to
  the galaxy in Mpc, the average inclination angle, $A_{{\rm HI}}$
   and $\alpha$ respectively.} 
\end{table}

\begin{figure}
\begin{center}
\epsfig{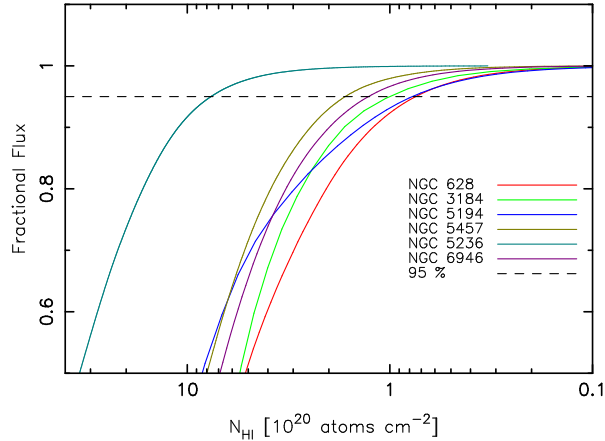}
\end{center}
\caption{The  fractional \HI 21-cm flux  as a function of  $N_{\rm
    HI}$. The  dashed horizontal line indicates the 95\% flux level.} 
\label{fig:cumulative}
\end{figure}
The only other quantity that we still  have to estimate from the data is
 $\Omega_{g}$.  To estimate $\Omega_{g}$,   we calculate  the flux of
the 
galaxy's 21-cm emission using only  the pixels where the \HI column
density exceeds a threshold value   $N_{\rm HI}$.  We find that the flux initially
increases as the threshold value  $N_{\rm HI}$  is lowered,  and then
saturates by $N_{\rm HI} = 
10^{19} \, {\rm cm}^{-2}$. The flux  does not increase much  if $N_{\rm
  HI}$ is lowered  further.   We use this to define the galaxy's
fractional flux  which is shown as a function of $N_{\rm HI}$ in 
Figure~\ref{fig:cumulative}.  The cut-off column density
$N_{\rm HI}^{c}$  is chosen such that   it encloses $ 95\%$ of the
galaxy's \HI 21-cm  flux, and we use this value of the \HI column
density to estimate $\Omega_{g}$.   Table~\ref{tab:nhic} gives the
values of $N_{\rm   HI}^{c}$ and $\Omega_{g}$ for the $6$   galaxies
that we have analyzed. Table~\ref{tab:nhic} also shows how
$\Omega_{g}$ changes if we use the $ 90 \%$ flux level instead of the  $95
\%$  level, clearly the values are similar   with a nearly  uniform
scaling  factor of $1.2$ for all $6$ galaxies.

\begin{table}
\centering
\begin{tabular}{l  c  c |  c}
\hline 
 & $N_{\rm HI}^{c}$ &  $\Omega_{g}$ &\\
Galaxy  & (atoms cm $^{-2}$) & (sterad) &
$\frac{\Omega^{[95\%]}_{g}}{\Omega^{[90\%]}_{g}}$ \\ 
& $\times 10^{20}$ & $\times
10^{-5}$ & \\
\hline \hline
NGC~628   & $0.75$ & $2.39$ & $1.24$ \\
NGC~3184  & $1.01$ & $0.69$ & $1.20$ \\
NGC~5194  & $0.78$ & $0.96$ & $1.33$ \\
NGC~5457  & $1.66$ & $4.83$ & $1.19$ \\
NGC~5236  & $7.64$ & $3.65$ & $1.18$ \\
NGC~6946  & $1.28$ & $2.65$ & $1.20$ \\
\hline
\end{tabular}
\label{tab:nhic}
\caption{Columns (2) and (3) show $N_{\rm HI}^{c}$ and   $\Omega_{g}$
  respectively, both have been defined using the  $95\%$  flux level. 
Column (4) shows how $\Omega_{g}$ changes if we use the $90\%$ flux level
instead.} 
\end{table}

\begin{table}
\centering
\begin{tabular}{l r r r r r r r}
\hline
Galaxy  & $\langle {\rm N}_{\rm HI} \rangle$ &  $A_{{\rm N}_{\rm HI}}$ &
$\alpha$ & kmin & kmax \\
& ($10^{20}$ cm$^{-2})$  &   & &
({\rm kpc}$^{-1}$) & ({\rm kpc}$^{-1}$) \\
\hline \hline
NGC 628  &  3.6 & $1.4\pm0.2$   &	$-1.6\pm0.1$    &  0.86	&  8.61 \\
NGC 3184 &  4.3 & $0.45\pm0.07$ &	$-1.3\pm0.1$    &  0.40	&  3.96 \\
NGC 5194 &  4.9 & $7.7\pm0.6$   &	$-1.7\pm0.1$    &  0.79	&  6.28 \\
NGC 5457 &  6.4 & $13.9\pm0.7$  &	$-2.1\pm0.1$   &  0.51	& 10.19 \\
NGC 5236 & 27.7 & $1.5\pm0.3$   &       $-1.8\pm0.1$    &  0.84	&  8.38 \\
NGC 6946 &  5.3 & $0.35\pm0.06$ &	$-1.7\pm0.1$    &  1.60	& 10.65 \\
\hline
\end{tabular}
\label{tab:result}
\caption{Columns (2), (3)  and (4) respectively show {\bf $\langle {\rm
    N}_{\rm HI} \rangle$  the mean \HI column density,  $A_{{\rm
      N}_{\rm HI}}$  the amplitude of the \HI   column density power
  spectrum at $k= 1 \, {\rm   kpc}^{-1}$  and the slope $\alpha$.}
  Columns (5) and (6)   show the $k$ range that was used to  
  determine the best-fit power law.}
\end{table}

We have used use  eqn.~(\ref{eq:ampform}) with the values given in
Tables~1 and 2  to calculate $ A_{{\rm N}_{\rm HI}}$ {\bf which is 
the amplitude of the \HI   column density power
  spectrum at $k= 1 \, {\rm   kpc}^{-1}$.} 
The  values are  listed  in Column (3) of Table~\ref{tab:result}. 
The $1\sigma$ errors have been  obtained by propagating the error from
$A_{{\rm HI}}$. We have  assumed that there is no error in estimating
$\Omega_{g}$ and the distance to the galaxy. {\bf For completeness, we have
shown   $\langle {\rm    N}_{\rm HI} \rangle$  the mean \HI column
density  in Column(2) and the slope $\alpha$ 
  in Column (4),  and present $k_{min}$ 
and $k_{max}$ which are the limits of the power-law fits in Columns
(5) and (6) respectively.} The $k_{min}$ and $k_{max}$ values have been 
calculated by applying the scaling factor $2 \pi/r$ to $U_{min}$ and
$U_{max}$  from \citet{2013NewA...19...89D}.

\begin{figure}
\begin{center}
\epsfig{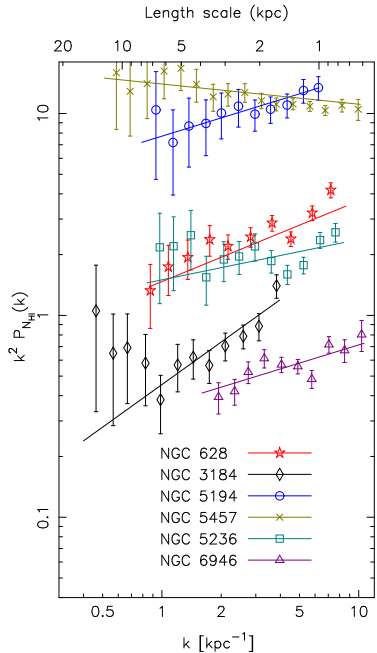}
\end{center}
\caption{Value of the $k^{2}P_{HI}(k)$ of the six galaxies in our
  sample are plotted against the wave number $k$. Note that all the
  galaxies expect NGC~5457 shows a positive slopes. Error bars in this
plot are obtained propagating the measurements by
\citet{2013NewA...19...89D}. Top margin gives the length scale ranges
corresponding to a $k$ value as $2~\pi/k$.}
\label{fig:ps}
\end{figure}

Figure~\ref{fig:ps} shows  $k^{2}\, P_{\rm HI}(k)$ as a function of
$k$ for  the six galaxies that we have analyzed. The top margin shows
the length scale corresponding to the $k$ values shown on the lower
margin.  For each galaxy the
data points and the best-fit power law have been shown only over the
$k$ range where a power law fit is possible.  
For the column density  $N_{\rm   HI}$ 
expressed in units of $10^{20} \, {\rm cm}^{-2}$, we find that the
measured $k^{2}\, P_{\rm HI}(k)$ has values in the range $\sim 0.03$ to
$\sim 20$. This indicates that the rms. fluctuations in 
$(N_{\rm   HI}/10^{20} \, {\rm cm}^{-2})$  are of the  order of unity
across the 
length-scales of approximately $1$ to $10 \, {\rm kpc}$ in all the $6$
galaxies that  we have analyzed. Further, in all the galaxies barring 
NGC 5457, the fluctuations go down with increasing length-scale. This
behavior  is reversed in  NGC 5457. 

\begin{figure}
\begin{center}
\epsfig{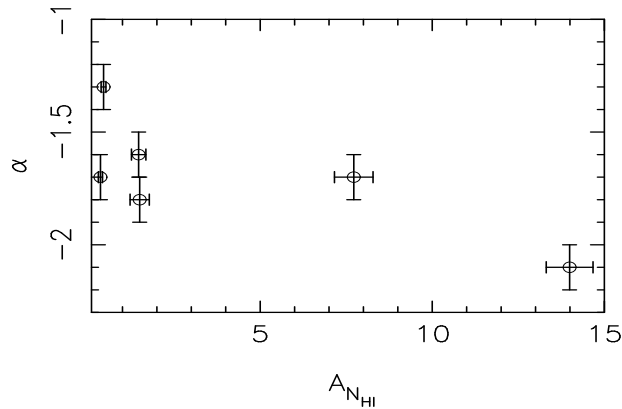}
\end{center}
\caption{The power law slope $\alpha$   plotted against  $A_{{\rm
      N}_{\rm HI}}$ which is the   amplitude of the $\NH$ power
  spectrum  {\bf  at $k= 1 \, {\rm   kpc}^{-1}$.}. }
\label{fig:A-alp}
\end{figure}

We have investigated if the  power law amplitude  $A_{{\rm N}_{\rm HI}}$ and
exponent $\alpha$ are correlated (Figure~\ref{fig:A-alp}). 
Though
visually there appears to be a  negative correlation,  
this is  mainly  because of the
rightmost point  which corresponds to   NGC~5457.  There
is no correlation if this data point is excluded. 

\section{Discussions and Conclusion}
We have restricted our analysis to   galaxies that are  nearly face on
with a maximum average inclination angle of $35^{\circ}$.  We
therefore do not expect geometrical effects to be very important, and
it is meaningful to compare  the results across  different galaxies in
our sample. Further, the   maximum $\NH$ fluctuation that we encounter
in  our analysis is  $\sim 4 \times 10^{20} \, {\rm cm}^{-2}$  which 
arises in the 
galaxy NGC~5457. This  is adequately  small  to justify  the  assumption
of low optical depth which has been adopted throughout our analysis. 

\begin{figure}
\begin{center}
\epsfig{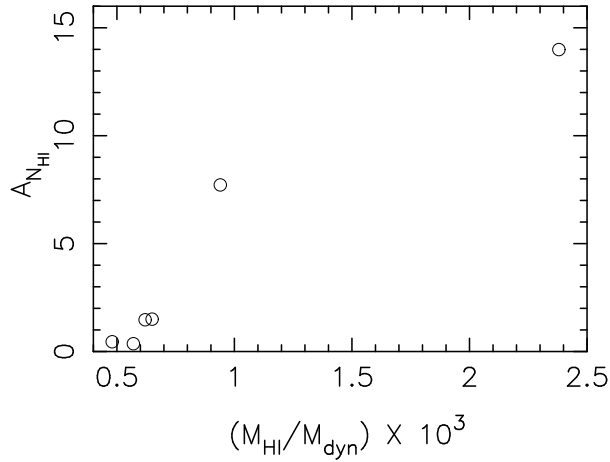}
\end{center}
\caption{The power spectrum amplitude $A_{{\rm N}_{\rm HI}}$ is
  plotted against the   \HI mass  fraction.  It is
  clear that these two   are highly correlated, the Spearman
  correlation coefficient has a value of  $0.94$. Data for the
  $M_{HI}$ and $M_{dyn}$ can be found   in
  \citet{2013NewA...19...89D}.} 
\label{fig:corr}
\end{figure}

The galaxy's radial \HI profile as well as the  substructure within the
galaxy's \HI disk both contribute to fluctuations in the \HI column
density.  {\bf Earlier work (e.g.   \citealt{2013NewA...19...89D}.)
 modeled  the galaxy's radial \HI profile through a window
 function, and the  analysis was carried out with the idea that
 one might, in future, be able to quantitatively separate the radial
 profile from the statistical fluctuations in the \HI disk. This
 idea has been abandoned in the current work. However,} 
we expect the galaxies to have a slowly
varying  radial profile  which falls off gradually away from the
center. \citet{2009MNRAS.398..887D}  have used simulations to quantify 
the effect of the radial profile on  the estimated  \HI power
spectrum. They find  that this effect is important  only at  
large length-scales which are comparable  to the extent of the
galaxy's \HI disk. The $k$ range where the galaxy's radial profile is
expected to be important has been excluded, and the analysis is
restricted to length-scales where   the 
measured \HI column density power spectrum  may be interpreted 
 as being entirely due to {\bf statistical fluctuations } in the galaxy's \HI
 disk.  The power law power spectrum indicates the presence of
 scale-invariant substructures  spanning length-scales $1$ to $10 \,
 {\rm kpc}$,   
 possibly originating from 2D compressible turbulence in
 the plane of  the galaxy's disk. There is evidence
 \citep{2008MNRAS.384L..34D} that this power law behavior does not
 extend to  length-scales much smaller than $1 \, {\rm pc}$. 
There is a break in the power law
 behavior at length-scales comparable to the thickness of the
 galaxy's \HI disk, and we have  a steeper power law which is usually 
 interpreted as 3D turbulence at small length-scales $(<  500 \, {\rm
   pc})$. 

It is quite plausible that the two different power laws
seen at large and small length-scales respectively are the
outcome of the same physical  process operating in the ISM of
galaxies.  The scale invariant behavior is interpreted as a signature
of turbulence  which is expected to be operational in the ISM. 
 It is now widely held that the small scale substructure is  
generated by  supernovae shocks whose  energy cascades down
to smaller  length-scales. However, it is unlikely that this can 
explain the observed kpc  scale structures.  We do not, at present,
have a clear  understanding 
of what  generates and  maintains these kpc scale structures in the
ISM. \citet{2013NewA...19...89D} have  studied if there is any
correlation between the slope of the power spectrum $P_{\rm HI}$ and
a variety of   physical and  dynamical parameters of the galaxies.
The presence of such a correlation is expected to provide some clue to
the physical mechanism  However,
the slope was not found to be correlated to any of the parameters like
the star formation rate, the velocity dispersion, the dynamical mass
and the \HI mass.   

The present work gives us access to $A_{{\rm N}_{\rm HI}}$ the
amplitude of the \HI column density power spectrum. {\bf The value of
$\sqrt{A_{{\rm N}_{\rm HI}}}$ gives an estimate of the rms. \HI column
density fluctuation at the wave number $k= \, 1 \, {\rm kpc}^{-1}$.
We find that this value varies by a factor of $6$ across the galaxies
that we have analyzed.} Like the slope, we
find that the amplitude  does  not have a correlation with   any of the
parameters mentioned earlier.  The amplitude is also found to be
uncorrelated to the inclination angle,  {\bf the mean \HI column
density $\langle {\rm    N}_{\rm HI} \rangle$}  and the value of
$N^c_{\rm   HI}$.  However, the amplitude $A_{{\rm N}_{\rm 
    HI}}$  appears to be correlated to the ratio of the galaxy's \HI
mass  to its dynamical mass  $M_{{\rm HI}}/M_{dyn}$ which gives  the \HI
mass fraction (Figure~\ref{fig:corr}).  The Spearman correlation
coefficient has a value  $0.94$ indicating a definite 
correlation.  The amplitude $A_{{\rm N}_{\rm HI}}$ increases with the
\HI mass fraction, however the data at present is insufficient to 
fit an algebraic relation between these two quantities. Our results
indicate that the underlying physical process is more effective in the
gas rich galaxies. Note however that care is needed in interpreting
this result as the galaxies have different values of the slope $\alpha$.  

The observed amplitudes and slopes of the \HI column
density power spectrum reported in this {\em Letter} provide an
important probe   of  the CDM  structure formation
paradigm. Cosmological simulations of galaxy 
formation incorporate a wide variety of physical process which are
relevant for shaping the properties of the ISM, and  there has  been
considerable    work  comparing the simulations with observations 
(eg. \citealt{2012ApJ...761...54E}). {\bf  This has, however, till now
  been mainly restricted to the one point statistics  $f(N_{HI},z)$. } 
  The  \HI column density power spectrum  provides a new 
observational handle for refining our understanding of the ISM and
comparing against simulations.

\section*{Acknowledgments}
We are indebted to Fabian Walter for providing us with the calibrated
\HI data from the THINGS survey. PD would like to thank Suman Majumdar
and Abhik Ghosh for  useful discussions.   
\bibliographystyle{mn2e}
\bibliography{references}

\end{document}